# Design of wavelength division multiplexing devices based on tunable edge states of valley photonic crystals

YuHui Han[a,b], HongMing Fei[a,b,*], Han Lin[c,*], MingDa Zhang[a,b], Xin Liu [a,b], XiaoRong Wang [a,b], BinZhao Cao [a,b], YiBiao Yang[a,b], LianTuan Xiao[a,b]

[a] *Department of Physics and Optoelectronics, Taiyuan University of Technology, Taiyuan 030024, China*

[b] *Key Laboratory of Advanced Transducers and Intelligent Control System, Ministry of Education, Taiyuan University of Technology, Taiyuan 030024, China*

[c] *Centre for Translational Atomaterials, Faculty of Science, Engineering and Technology, Swinburne University of Technology, Hawthorn, Victoria 3122, Australia*

*Corresponding author: feihongming@tyut.edu.cn; hanlin@swin.edu.au

**Abstract**

Wavelength division multiplexing (WDM) devices are key elements of Photonic integrated circuits (PICs). Conventional WDM devices based on silicon waveguides and photonic crystals have limited transmittance due to high loss introduced by the strong backward scattering from defects. In addition, it is challenging to reduce the footprint of those devices. Here we theoretically demonstrate a WDM device in the telecommunication range based on all-dielectric silicon topological valley photonic crystal (VPC) structures. We tune its effective refractive index by tuning the physical parameters of the lattice in the silicon substrate, which can continuously tune the working wavelength range of the topological edge states, which allows designing WDM devices with different channels. The WDM device has two channels (1470 nm-1523 nm and 1548 nm-1609 nm), with contrast ratios of 22.4 dB and 24.9 dB, respectively. The principle of manipulating the working bandwidth of the topological edge states can be generally applied in designing different integratable photonic devices, thus it will find broad applications.

**Key words**：Valley photonic crystal，Topological edge states，Wavelength division multiplexing, Photonic integrated circuit

# 1. Introduction

Due to the high energy efficiency, high speed and broad bandwidth, all-optical communication and quantum computing have broad potential applications. Photonic integrated circuits (PICs) are the platform for all-optical communication and quantum computing. Therefore, it is desired to develop PICs composed of integrated functional devices, such as light sources [1], optical modulators [2] and transceivers [3], wavelength division multiplexing (WDM) devices [4, 5] and photodetectors [6]. Among them a WDM device is one of the key elements due to the capability of significantly enhancing the processing speed and broadening the bandwidth by simultaneously using multiple working bands.

A WDM system have a multiplexer which combines different optical signal at different wavelengths into one channel, and a demultiplexer that transmits optical signal into different channels according to the wavelength. Therefore, an ideal WDM device should have high transmittance to minimize the loss and high contrast ratio between different channels at each working wavelength. There are several silicon photonic WDM devices have been proposed, such as the cascaded Mach-Zehnder interferometry based on the modal coupling and interference of the waveguide modes[5], and the arrayed waveguide grating structure based on the Rowland circle principle[7]. However, those devices inevitably require a millimeter-level large size to ensure good phase changes for effective interference, which has become a bottleneck of development of miniaturized silicon PICs. In comparison, the footprint of WDM devices can be further reduced by using photonic crystal (PC) structures[8-14] based on the photonic bandgap and optical localization of light in defect lines. More importantly, for all those silicon photonic WDM devices, including conventional and PC devices, the inevitable backscattering introduced by the manufacturing defects decreases the forward transmittance, which is undesired.

The recent development of topological photonic crystals (TPCs) [15-19] can realize robust unidirectional and anti-scattering transmission of light waves. Valley photonic crystals (VPCs) [20-30] are a kind of TPCs, which achieve high forward transmittance based on the spin-valley locking effect. Compared to other TPCs, VPCs generally have larger working bandwidth and can be fabricated using conventional dielectric material, such as silicon. The recent experimental demonstration of VPCs working in telecommunication wavelength region shows great potential of applying VPCs in silicon PICs [25]. It has been theoretically demonstrated a dual-band WDM devices by using the high-order bandgaps in VPCs [31]. However, theoretical design requires metallic materials working in microwave region, which cannot be used in silicon PICs. Therefore, there is no demonstration of WDM devices based on VPCs in the telecommunication wavelength region, which are highly desired.

In this work, we theoretically demonstrate a WDM device (Figure 1) based on silicon VPCs, which is achieved by tuning the bandwidth of topological edge states. The bandwidth of the topological edge states is tuned by controlling the effective refractive index of the lattices at the boundary, which can be realized by adjusting either the refractive index of the material or size of lattices. The resulted WDM device has two broadband channels (1470 nm-1523 nm and 1548 nm-1609 nm) in the telecommunication region, with high contrast ratios of 22.7 dB and 24.9 dB, respectively, which are much higher than conventional silicon photonic WDM devices [8]. In addition, the designed device can be fabricated by current mature complementary metal-oxide-semiconductor (CMOS) technology. The principle of manipulating the topological edge states can be generally applied to design different functional photonic devices, such as optical delay lines, optical modulators and interferometers. Therefore, our work opens new possibility in designing silicon photonic devices based on VPC structures with microscale footprint for PICs, and will find broad applications.

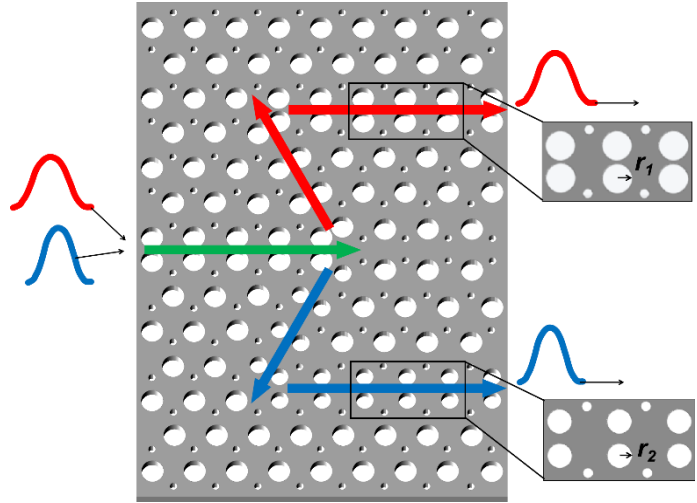

Figure 1 3D Schematic of a WDM device based on tunable edge states in a VPC structure.

## 2. Design of valley photonic crystal

The 3D schematic of the WDM device based on the tunable edge states in a VPC structure is shown in Figure 1, in which incident light at different wavelengths (indicated by the red and blue curves) propagates through different channels. In this design, the tuning of edge states is realized by tuning the radius ($r_1$ and $r_2$) of the lattices at the boundary. In order to design the WDM device, it is necessary to first design a VPC with a large topological photonic bandgap, which defines the overall working bandwidth of the device. The honeycomb lattice is a commonly used PC structure for designing VPC structure due to the Dirac point in the photonic band structure[30]. Figure 2(a) shows the unit cell of a silicon ($n_{Si}$ = 3.47) honeycomb photonic crystal structure with low index (air) lattices. The lattices constant is $a$ = 450 nm, and the thickness is $h$ = 220 nm. Initially, the $r_a$ is equal to $r_b$ (100 nm), so the structure has $C_{6V}$ rotational symmetry, which results in a Dirac point in the band structure (Figure 2(b)). Here we use the commercial 3D time-domain difference finite element (FDTD) software (Lumerical FDTD Solutions) to calculate the band structure. Then the $C_{6V}$ symmetry is reduced to $C_{3V}$ symmetry by simultaneously increasing $r_a$ (to 120 nm) and decreasing $r_b$ (to 40 nm) to introduce a topological photonic bandgap (1434 nm-1687 nm) due to the degeneracy of the K and K' valleys, indicated by the black lines in Figure 2 (b). In this way, we get VPC1 based on Valley Hall Effect[22] (see Supplementary

Section S1 for details). The VPC2 is a mirror image of VPC1, which shares the same bandgap.

The topological property of VPC1 and VPC2 can be described by the valley-dependent topological index [19,20]:

$$C_{\tau z} = \frac{\tau_z \, sgn(\lambda_{\varepsilon r}^P)}{2} \tag{1}$$

which are $C_K = 1/2$ and $C_{K'} = -1/2$, for K and K' valleys, respectively. Therefore, even the total Chern number is equal to zero, we can define this new non-zero topological invariant, namely the valley Chern number [20]: $C_V = C_K - C_{K'}$, which is 1 for VPC1 and -1 for VPC2, respectively. The opposite valley Chern numbers are also a necessary condition for topological edge states to be generated at the boundary between VPC1 and VPC2. The photonic band structure of the beard-shape boundary (Figure 2(c)) between VPC1 and VPC2 is shown in Figure 2(d), in which the topological edge state is indicated by the blue curve.

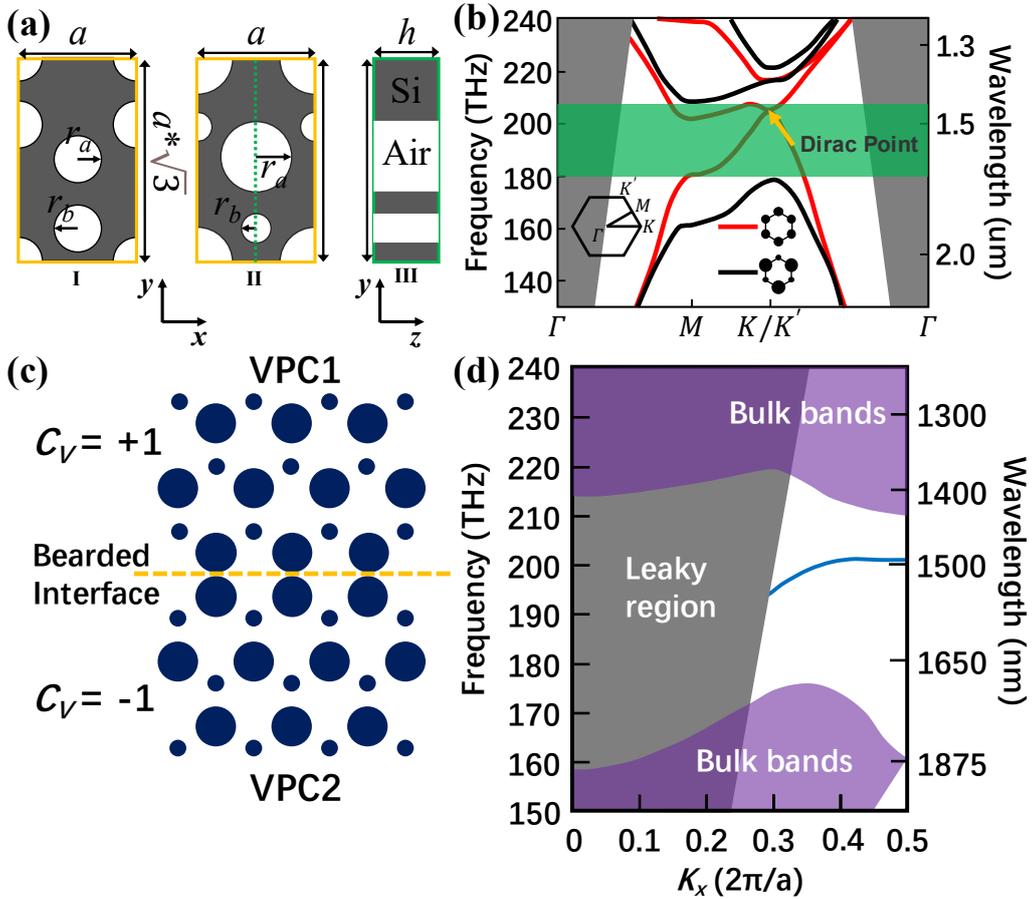

Figure 2 (a) Schematic diagrams of photonic crystal structures. The yellow rectangular marks the unit cell of the structure, and the green rectangular shows the cross-sectional view at the green dashed line. The lattice constant is *a*, and the thickness of the structure is *h*. The gray area represents the Si material, the white area represents the low index cylinder. (b) The photonic band structure of the honeycomb (red curves) and VPC1 (black curves). The yellow arrow marks the Dirac point. The green shadow marks the TE mode band gap, and the gray area is the air light cone. (c) Schematic of the beard-shape boundary, which is composed of mirror-symmetrical VPC1 and VPC2, respectively. (d) Beard-shape boundary energy band diagram, the blue curve shows the topological edge state, the purple area represents the bulk states, and the gray area is the air light cone.

There are two possible types of the boundaries, namely the zigzag-shape and beard-shape boundaries to achieve topological edge states. The calculated edge states of the two boundaries are shown in Figure 3(a), indicated by the red and blue curves, respectively. The edge state of the zigzag-shape boundary connects the top and bottom of the bandgap, which is beneficial to achieve broadband transmission of VPC devices[29]. In comparison, bandwidth (1470 nm – 1523 nm) of the edge state of the beard-shape boundary is narrower than the bandgap (blue curve in Figure 3(a)), which allows tuning the position of the working bands within the bandgap. It can also be seen from the transmittance spectra in Figure 3(b) that the bandwidth of the zigzag-shape boundary is wider than that of the beard-shape boundary. In this study, as it is required to achieve different working bands within the bandgap, beard-shape boundary is used.

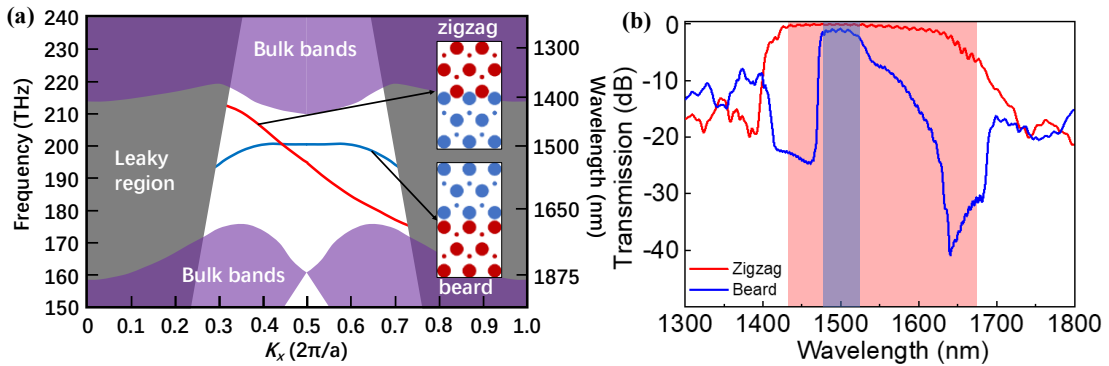

Figure 3 (a) Band diagram of edge states of zigzag-shape and beard-shape boundaries. The schematics of the boundaries are shown in the insets; (b) Transmittance spectra of zigzag-shape and beard-shape boundaries.

## 3. Manipulation of working bands of topological edge states

In this study, two approaches are applied to manipulate the working bands (frequency ranges) of the topological edge states, namely tuning the size or the refractive index of the low index lattices at the boundary. The radius of the low index lattices is $r_e$ (Figure 4(a)), which is gradually tuned from 120 nm to 90 nm. Along with the decrease of $r_e$, the working wavelength redshifts due to the increase of the effective refractive index as shown in Figure 4(c). In the meantime, some trivial modes show up indicated by the dashed lines in Figure 4(c), which cannot support scattering-free unidirectional transmission[32] (see Supplementary Section S2 for details). Similar effect can be realized by tuning the refractive index of the low index material $n_e$ without changing the $r_e$. Here $n_e$ is changed from 1.0 to 1.9 (Figure 4(b)). The resulted frequency shift of edge states is plotted in Figure 4(d), which also redshifts due to increase of effective refractive index. As shown in Figures 4(e) and (f), both the tuning of radius and the refractive index can continuously shift the working band within a large wavelength range (>150 nm). In addition, the bandwidth remains relatively the same during the tuning process, which allows choosing multiple channels as long as the working bandwidths of the channels do not overlap. It should be noted that this method of tuning the frequency range of edge states can be generally applied to different VPC structures (more detail information can be found in Supplementary Section S3).

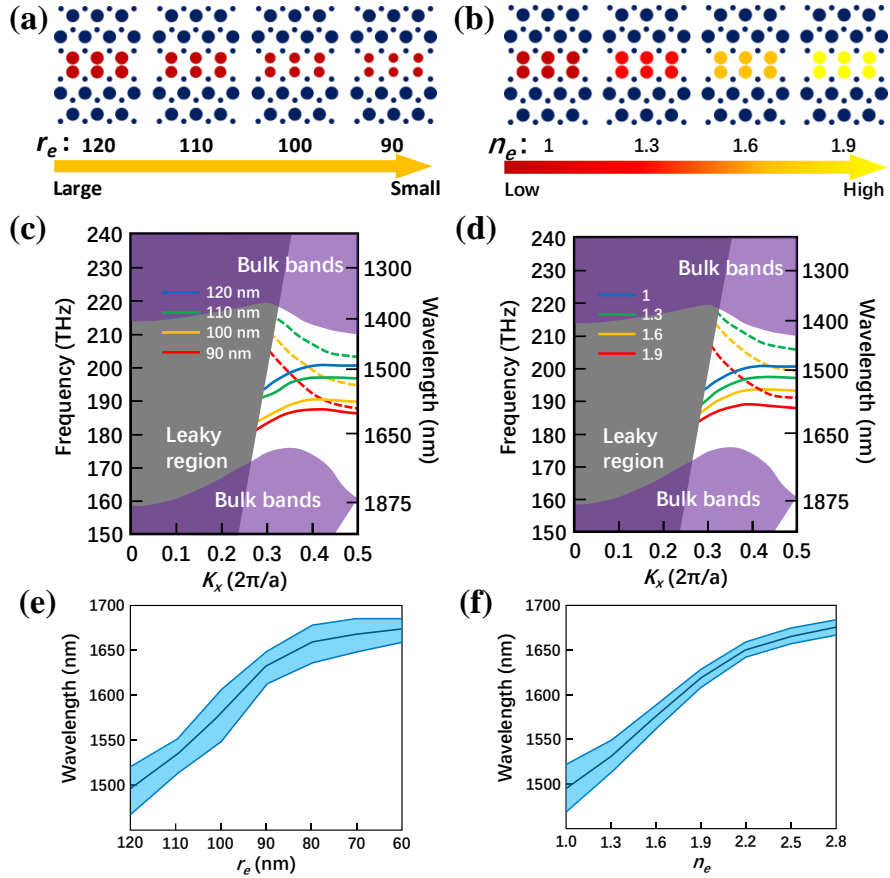

Figure 4 Tuning the frequency range of edge states. (a) and (b) are the schematic diagrams of the change of the radius $r_e$ and the refractive index $n_e$ of the rows of low index lattices at the boundary. (c) and (d) photonic band diagrams of edge states with different $r_e$ and $n_e$. The solid and dashed color lines represent the topological and trivial edge states, respectively. (e) and (f) show the change in the frequency range of the edge states with $r_e$ and $n_e$, respectively. The solid lines mark the central wavelength of the frequency range.

In order to visualize the modal field distribution, the intensity of the electric field ($|E|^2$) of edge states is plotted in Figure 5. One can see that the electric field is mainly confined within the high index material (silicon). Therefore, when $r_e$ is reduced, the silicon area becomes larger (Figures 5(a) and (b)), resulting in a high effective refractive index that can support edge states at longer wavelength. Similar effect can be identified from in Figures 5(c) and (d), in which $n_e$ increases from 1 to 1.8. As shown in Figures 4(e) and (f), this method is able to tune the frequency range of edge states with very fine step ($\Delta r_e = 10\ nm$ or $\Delta n_e = 0.3$), which can generally be applied to control the working band of any VPC devices. However, the WDM devices require well separate working bands. Therefore, it is necessary to choose $r_e$ or $n_e$ with large intervals to ensure no overlapping of working bands of different edge states.

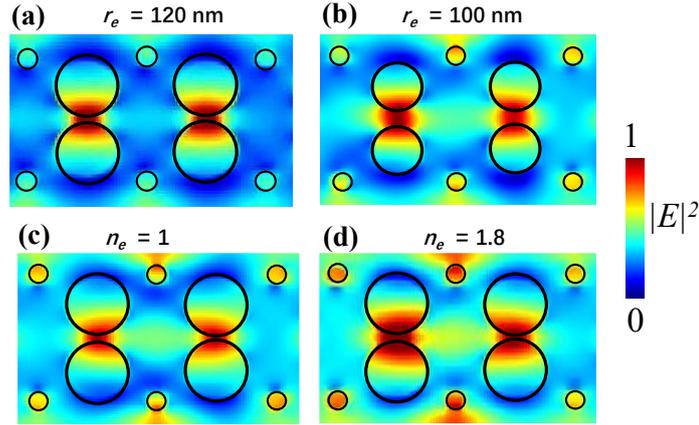

Fig. 5 Edge state modal field intensity distributions. (a) and (b) correspond to the radii of lattices at the boundary of 120 nm and 100 nm, respectively, (c) and (e) correspond to the refractive index of the lattices at the boundary of 1 and 1.8, respectively.

## 4. Design of WDM devices

The schematic of the WDM device with two channels composed of VPC1 and VPC2 is shown in Figure 6(a). There are four waveguides in the devices, namely WG1, WG2, WG3 and WG4, in which WG4 is the input waveguide, and WG1 and WG3 are the two working channels. The $r_e$ is 120 nm for WG1 and 100 nm for WG3, respectively. The photonic band diagram of the edge states of WG1 to WG4 is shown in Figure 6(b), in which the working band for WG1 and WG3 are band A (1470 nm-1523 nm) and band B (1548 nm-1609 nm), respectively. This design enables the incident light at different wavelengths to be transmitted along the different waveguides to achieve the goal of wavelength demultiplexing. Due to the inverse combination of VPC1 and VPC2 in WG2 compared to WG1 and WG3, the polarization of the edge state is different from the ones in WG1 and WG3. This is also confirmed by the opposite group velocity shown in Figure 6(b). Therefore, the incident light from WG4 cannot propagate along WG2. The working bands for WG1 (band A) and WG3 (band B) are marked by the red and green stripes in Figure 6(b), which are well separated. Here the WG4 is designed to be exactly the same as WG1. Therefore, the incident light within band A travels from WG4 to WG1. Meanwhile, for the incident light within band B, it is in the scattering area in the topological mode of WG4 and WG1. Therefore, strong out-of-plane scattering prevents light propagating WG4 to WG1. As a result, the incident light within band B transmits through WG3. In this way, broadband incidence is demultiplexed according to the wavelength.

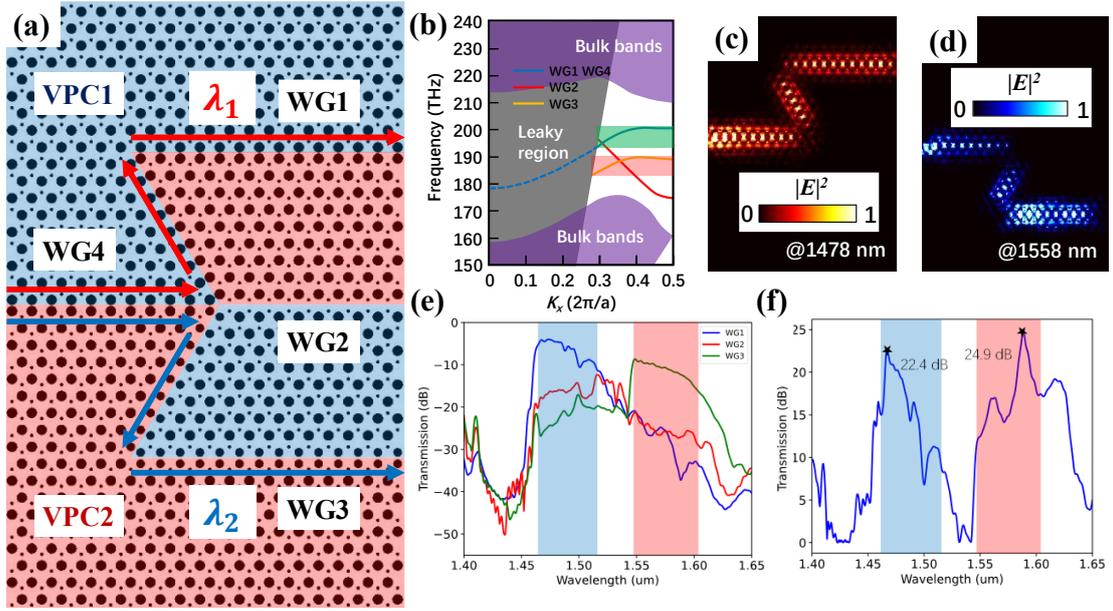

Figure 6 (a) Schematic of the WDM device, where the red and blue arrows mark the propagation path for incident light in different bands. The blue and red areas indicate VPC1 and VPC2, respectively. (b) Photonic band diagram of the edge states of WG1 to WG4, the green and red areas mark band A and band B, respectively. (c) and (d) are the electric field intensity distributions of incident light at the wavelengths of 1478 nm and 1558 nm, respectively. (e) Transmittance spectra of WG1, WG2 and WG3. (f) Transmittance contrast ratio plots of WG1 and WG3, respectively. The blue and red areas mark the band A and band B, respectively.

The electric field intensity distributions are shown in Figures 6(c) and (d), respectively. As one can see the incident light can transmit through different channels according to the wavelength with high contrast. The transmittance spectra of different waveguides are plotted in Figure 6(e), which show high transmittance in the working band of the waveguides due to robust unidirectional propagation properties of the topological waveguides. Meanwhile, the transmittance of WG2 is low in both working bands. The other key parameter to characterize the performance of the WDM device is the transmittance contrast (TC) of different channels, which can be calculated as TC = $|T_{WG1} - T_{WG3}|$, where TC is the transmittance contrast, $T_{WG1}$ and $T_{WG3}$ are the transmittances of WG1 and WG3, respectively. The transmittance contrast curve is shown in Figure 6(f), which show a high contrast ratio up to 22.4 dB and 24.9 dB in band A and band B, respectively. The contrast ratio is higher than most of silicon WDM devices [8]. Therefore, this WDM device achieves high performance wavelength demultiplexing function in the telecommunication region.

## 5. Conclusion

In conclusion, we have demonstrated a high performance WDM device based on tuning the frequency range of the edge states. The frequency range is controlled by manipulating the effective refractive index of the VPC waveguide structure, which is realized by varying the radius or refractive index of the low index lattices at the boundary. The designed WDM device has two channels at the wavelength regions of 1470-1523 nm and 1548-1609 nm, respectively. The transmittance contrast of the two channels can be as high as 22.4 dB and 24.9 dB. This design is based on all-dielectric silicon material working in the telecommunication wavelength range, which can be fabricated by current CMOS technology and is suitable for highly integrated and compact photonic chip. More importantly, the tuning principle can be generally applied to design different VPC devices. Therefore, this work opens new applications of VPC structures and will find broad applications, especially in PICs.

## Acknowledgment

This work was sponsored by the Young Scientists Fund of the National Natural Science Foundation of China (Grant No.11904255), the Key R&D Program of Shanxi Province (International Cooperation), China (Grant No.201903D421052), the Applied Based Research Program of Shanxi Province (Youth Fund), China (Grant No.201901D211070).